\documentclass[journal]{IEEEtran}
\usepackage{colortbl}
\usepackage{color}
\usepackage{multirow}
\usepackage{cite}
\usepackage{balance}
\usepackage{tipa}
\usepackage{amsmath}
\usepackage{enumitem}
\usepackage{comment}
\usepackage{makecell}
\usepackage{lipsum}
\usepackage{multicol}
\usepackage{algorithmic}
\usepackage[normalem]{ulem}
\setlist[itemize]{noitemsep, topsep=0pt}

\usepackage{lipsum}
\ifCLASSOPTIONcompsoc
    \usepackage[caption=false, font=normalsize, labelfont=sf, textfont=sf]{subfig}
\else
\usepackage[caption=false, font=footnotesize]{subfig}
\fi

\usepackage{pdfsync}
\usepackage{amssymb}
\usepackage{bbold}
\usepackage{color}
\usepackage{physics}
\usepackage[ruled,vlined]{algorithm2e}

 \usepackage[font=small]{caption}

\ifCLASSINFOpdf
\usepackage[pdftex]{graphicx}
\DeclareGraphicsExtensions{.pdf,.jpeg,.png,.jpg}
\else
\usepackage[dvips]{graphicx}

\DeclareGraphicsExtensions{.eps}
\fi
\graphicspath{{figs/}}
\usepackage{epstopdf}
\usepackage{booktabs}
\usepackage{amsmath}

\begin{document}

\title{Extremum-Seeking Adaptive-Droop for Model-free and Localized Volt-VAR Optimization}

\author{Hongda Ren,~\IEEEmembership{Student Member,~IEEE,}
        Rahul Ranjan Jha,~\IEEEmembership{Student Member,~IEEE,}
		Anamika Dubey,~\IEEEmembership{Member,~IEEE,}
        Noel N. Schulz,~\IEEEmembership{Fellow,~IEEE }
		\vspace{-0.4 cm}
         \thanks {This material is based upon work supported by the U.S. Department of Energy under Award Number DE-IA0000025. The views and opinions of authors expressed herein do not necessarily state or reflect those of the United States Government or any agency thereof.}  
         \thanks {H. Ren, R. Jha, A. Dubey, and N. N. Schulz  are with the School of Electrical Engineering and Computer Science, Washington State University, Pullman, WA, 99164 e-mail: hongda.ren@wsu.edu,  rahul.jha@wsu.edu, anamika.dubey@wsu.edu, noel.schulz@wsu.edu. }
	}

\maketitle
\begin{abstract}
In an active power distribution system, Volt-VAR optimization (VVO) methods are employed to achieve network-level objectives such as minimization of network power losses. The commonly used model-based centralized and distributed VVO algorithms perform poorly in the absence of a communication system and with model and measurement uncertainties. In this paper, we proposed a model-free local Volt-VAR control approach for network-level optimization that does not require communication with other decision-making agents. The proposed algorithm is based on extremum-seeking approach that uses only local measurements to minimize the network power losses. To prove that the proposed extremum-seeking controller converges to the optimum solution, we also derive mathematical conditions for which the loss minimization problem is convex with respect to the control variables. Local controllers pose stability concerns during highly variable scenarios. Thus, the proposed extremum-seeking controller is integrated with an adaptive-droop control module to provide a stable local control response. The proposed approach is validated using IEEE 4-bus and IEEE 123-bus systems and achieve the loss minimization objective while maintaining the voltage within the pre-specific limits even during highly variable DER generation scenarios.

\end{abstract}
	%
	\IEEEpeerreviewmaketitle

\section{Introduction}

\IEEEPARstart{E}{lectric} power distribution systems are rapidly transforming to active networks with the integration of controllable distributed energy resources (DERs). In this regard, the rapidly varying generation patterns as a result of DER integration that are leading to bus voltage violations and fluctuations and motivate the need for active voltage control. Traditional voltage control methods that employ legacy devices such as capacitor banks and voltage regulators are not effective in managing the grid voltages for systems with high-levels of DER penetrations. This calls for employing controllable smart inverters to dispatch/absorb reactive power \cite{osti_1166677,8332112}. Optimal control of smart inverters can not only help manage the grid voltages but also provide additional grid services such as loss minimization, conservation voltage reduction. The focus of this paper is on coordinating the smart inverters through Volt-VAR optimization (VVO) for loss minimization while maintaining the bus voltages within the acceptable limits. 

Centralized optimal power flow (OPF) algorithms have been proposed for smart inverter coordination with the objective of loss minimization, conservation voltage reduction, etc. \cite{Farivar,Su,bileveljha, chamana2018optimal}. However, the resulting nonlinear OPF problem is challenging to solve for a large unbalanced distribution system with numerous controllable agents \cite{7403309}. Furthermore, centralized methods require a network model and an extensive communication infrastructure to gather measurement and to dispatch decisions. Thus, they are susceptible to model and measurement uncertainty and vulnerable to communication failures. Lately, distributed Volt-VAR control (VVC) has emerged as a solution to some of the problems posed by centralized methods. They employ a multi-agent set up to solve the network-level optimization problem upon decomposing the centralized problem into multiple sub-problems that are assigned to different agents. The agents solve local optimization problems and exchange the boundary variables with the neighboring agents. The agents converge on a solution upon multiple rounds of exchange of information with the neighbors. Some of the common methods employed for solving distributed OPF include, Dantzig-Wolfe decomposition \cite{shi2016decentralized}, alternate direction method of multipliers (ADMM) \cite{robbins2013two}, auxiliary problem principle; see \cite{7990560} for a comprehensive review on distributed OPF algorithms. While the distributed methods have low communication requirements, it takes about 100s of iterations among the agents to converge \cite{7990560, 6502290, 8573893}. Thus, timely decision-making using the existing distributed methods requires a fast communication infrastructure that is typically not available. Further, the distributed methods also rely heavily on the network models and, thus, are susceptible to model and measurement uncertainties \cite{7990560}. 

Both centralized and distributed optimization solutions are not fast enough to manage rapidly changing phenomena, such as highly variable PV generation during cloud transients. Further, both require network models and communication infrastructure to gather measurements, and implement decisions. On the contrary, local Volt-VAR control, such as using droop-curve for smart inverter control, are model-free and only require local measurements for decision-making \cite{8332112}. The local droop-based control methods are known to pose instability issues \cite{8636257}. Lately, adaptive droop-control methods have been proposed to ensure stable local control under external disturbances \cite{singhal2018real}. However, local droop-based control methods are unable to coordinate the smart inverters to achieve a network-level objective such as loss minimization. Recent work uses the extremum seeking approach to locally control smart inverters and coordinate their receive power dispatch for CVR for a balanced distribution system~\cite{7350258}. The approach, however (1) requires a communication infrastructure to broadcast the feeder head measurement to all distributed inverters, (2) assigns specific perturbation frequencies and design specific filter to each controller to isolate the impacts of specific localized perturbations; the frequency bandwidth may limit the number of inverters, (3) directly controls reactive power dispatch and hence is not compatible with the IEEE 1547 standard. It also does not include further analysis and results for an unbalanced power distribution system. The approach proposed in this paper specifically mitigates these limitations for the local VVO methods. 

Another Volt-VAR control strategy includes the coordination of both centralized and local Volt-VAR controls methods. In this regard, recent methods include DER coordination by reactive power control with the aid of system model and communication. For example, in \cite{8600388}, an IEEE 1547 compatible rule-based local Volt-VAR control is proposed to mitigate voltage violations affected by the fluctuations of PV real power. The proposed algorithm uses a linear programming (LP) approach to compute the slope of VVC curves using the system model and communicates the updated inverter settings obtained upon solving the LP problem \cite{8600388}. In \cite{8003321}, authors use linear approximation of AC power flow and load forecasting methods to update VVC coefficients. Another coordinated control approach with central and local active and reactive power controls minimizes the total PV active power curtailment to maintain the voltages within the acceptable limits \cite{6755528}. In \cite{8610230}, authors tested the coordinated control of reactive power from DERs to provide voltage support for the bulk transmission system. The historical smart meter data was used to design optimal active power-reactive power characteristic Q(P) curves for local voltage control of PV inverters and  validated on a three-phase unbalance system \cite{7478138}.


\begin{table}[h]
    \centering
		\caption{Comparison of Volt-VAR controls }
		\label{tab:VVCcomparison}
\vspace{-0.2cm}
\begin{center}
    		\begin{tabular}{c|c| c| c }
    		\hline
    		\hline
    		\small
    		 {\hspace{-4pt}Methods} & Centralized & Distributed & Local\\
			\hline
			{\hspace{-4pt}Communication}& \makecell{all controllers\\ with center}  &  \makecell{data exchange \\among neighbors \\ } & \makecell{  no \\requirement}\\
			  \hline
            {\hspace{-4pt}Optimization}& optimal solution & \makecell{ close to \\optimal solution} & \makecell{no \\ optimization }\\
            \hline
            {\hspace{-4pt}System Info}& system model  & sub-area models & \makecell{no model \\ required} \\
            \hline
             {\hspace{-4pt}Computation}& high   & \makecell{ medium (parallel \\ computation)} & low \\
            \hline
            {\hspace{-4pt}Control Interval}& \makecell{computation\\ communication} & \makecell{ computation\\ communication}  &  real time  \\
            \hline
            {\hspace{-4pt}Measurements}& \makecell{P, Q, V, I, \\ tap \& cap status} & \makecell{ P, Q, V, I, \\ tap \& cap status\\in sub-areas }  & \makecell{ local\\ P, Q, V, I}\\
            \hline		
            \hline
			\end{tabular}
		\vspace{-0.3cm}
		\end{center}
\end{table}

Although several centralized, distributed, local and their combination optimization and control methods have been proposed in the related literature, there remain crucial limitations in smart inverter coordination for grid services that need addressing (see Table \ref{tab:VVCcomparison}). To sum up, the existing centralized and distributed VVO methods heavily rely on network models and communication infrastructure and are not fast enough to manage rapidly varying phenomena. The local control methods, while being model-free, are not able to achieve network-level optimization using only local measurements. Although few have applied local VVO algorithms to unbalanced power distribution systems~\cite{7478138}, further analysis is needed to understand the analytical properties of purely local VVO control approaches for unbalanced power systems and their ability to respond to fast-changing phenomena such as DER generation variability.  Additional analysis is also needed to make the local Volt-VAR control methods compatible with IEEE 1547 standard. To address the aforementioned gap in the existing literature, in this paper, we introduce a purely local measurement-based model-free VVO to control smart inverters for loss minimization in an unbalanced power distribution system. The proposed approach utilizes an extremum-seeking approach to optimize for the feeder losses using only local measurements while maintaining bus voltages within acceptable limits using an adaptive-droop control. The contributions of the proposed approach are detailed below:

\begin{itemize}[nolistsep,leftmargin=*]
\item {\em Communication and model-free local VVO:}  The proposed approach introduces localized optimization algorithms to reach model-free, and communication-free VVO for loss minimization. The extremum seeking algorithm drives the droop-based Volt-VAR controller to the network-level optimal solutions. The proposed method uses only local measurements and thus is not impacted by the control conflicts due to other local controllers. Local measurements also encourage modularity as the number of local ES controllers (at individual PV locations) is not limited by the perturbation frequency bandwidth. The proposed approach also works well with the legacy voltage control devices such as voltage regulators. The local design ensures the proper functionality of the controller with varying network models. 
\item {\em Adaptive and Stable Fast VVC:} The conventional droop-based Volt-VAR controller (based on IEEE 1547-2018 standard) is enhanced with adaptive control strategies to handle control instability resulting from high variance inputs. The proposed VVC design ensures fast response to the changes in loads, network topology, and intermittent solar generation, and legacy voltage regulation devices.
\item {\em Analytical results for Unbalanced Power Distribution System for Optimality:} 
We analyze the convex relationship between power loss and reactive power injection for the balanced and unbalanced power distribution systems, which builds the foundation for extremum-seeking converging to the optimal solutions. The analysis of the unbalanced  system  includes  mutual  coupling  and  unbalanced loading. 
\end{itemize}
 

\section{Background}
The objective of this paper is to employ local model-free and measurement-based VVO for loss minimization by controlling the reactive power dispatch from smart inverters. To this regard, we review extremum seeking algorithm that is a measurement-based local control approach used to seek the optimum of a function. For fast changing phenomenon, several extremum-seeking controllers may lead to instability. This calls for other control loops to stabilize the control response. Adaptive droop-control is effective in such cases. In what follows, we introduce extremum seeking optimization and droop-based VVC methods.

\begin{figure}[t]
\includegraphics[width=0.48\textwidth]{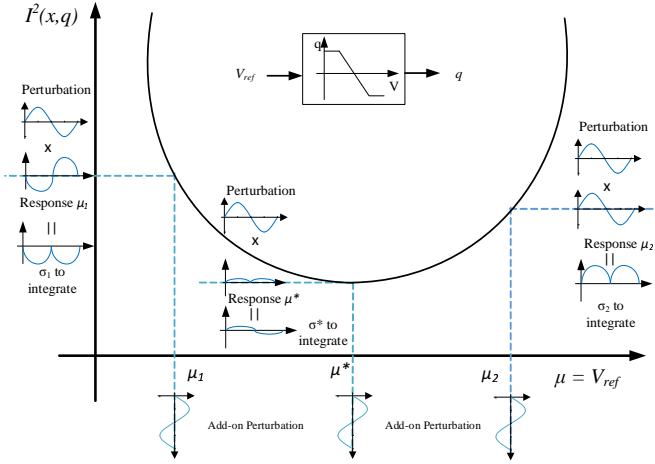}
	\caption{Extreme Seeking Scheme Concept \cite{KRSTIC2000595, brunton2019data}}
	\label{fig:ESschemeConcept}
	\vspace{-0.70cm}
\end{figure}

\begin{figure*}[!t]
 \centering
\includegraphics[width=0.96\textwidth]{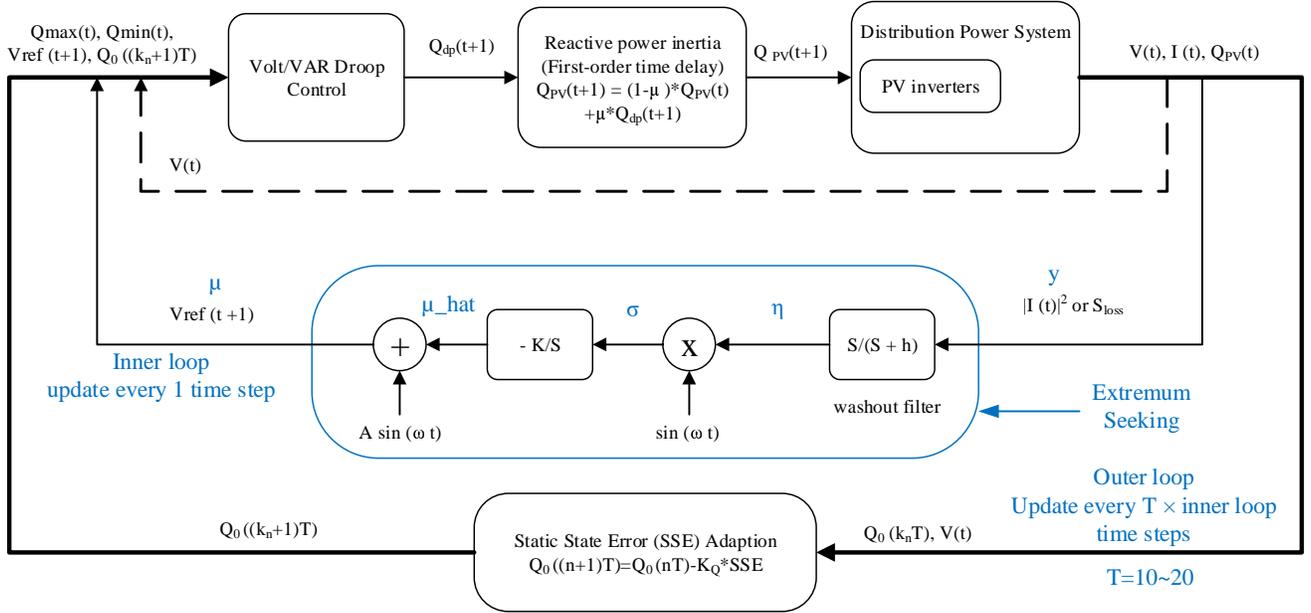}
\vspace{-0.1cm}
	\caption{Proposed approach for Localized Volt-VAR Optimization}
\vspace{-0.3cm}
	\label{fig:proposedvvo}
\end{figure*}

\subsection{Extremum Seeking Optimization (Preliminaries)}
Extremum seeking is a measurement-based real-time local control method that seeks optimum of a function without requiring the model parameters or communication with other nodes. The basic principle is to slightly perturb the system states and use the local measurements to obtain the gradient descent direction for the objective function \cite{KRSTIC2000595}. The flow chart and notations are highlighted in blue in Fig.~\ref{fig:proposedvvo}. A sinusoidal signal, $sin({\omega}t)$, is used as a time continuous perturbation to the physical system to estimate gradient of the objective function. 
The local measurements are used to estimate current value of the objective function, $y$, that needs to be optimized. The estimate of the objective function, $y$, is passed through a washout filter to remove noise having frequency component lower than $\omega_h$. The resulting signal $\eta$ is multiplied with the sinusoidal perturbation and passed through an integrator of gain $K$. This addition on the original perturbation leads $\hat{\mu}$ in the gradient direction to minimize objective function $y$. The algorithm converges to a local minimum when $K > 0$. Otherwise, it will converge to a local maximum. 

The mathematical expressions for the extremum seeking algorithm is  given in (\ref{eqES1})-(\ref{eqES3}) \cite{KRSTIC2000595}. 
\vspace{-0.05cm}
\begin{small}
\begin{eqnarray}\label{eqES1}
   \dot{y}  &=& \dot{\eta}-h\eta \\
   \label{eqES2}
    \dot{\hat{\mu}}&=&-K*Asin (\omega t)*\eta\\ 
    \label{eqES3}
    \mu&=&\hat{\mu}+Asin\omega t
    \end{eqnarray}
\end{small}
\vspace{-0.4cm}

Fig.~\ref{fig:ESschemeConcept} illustrates the working principle of the extremum seeking algorithm to estimate the gradient and reach local minimum point of an unknown function, $I^2$ for instance. $I^2$ is the function of $x$ and $q$ where, $x$ and $q$ are uncontrollable and controllable variables, respectively. Here, $\mu$ corresponds to the decision variable, $V_{ref}$, which has direct impact on $q$ through a linear droop function. Let, $\mu^*$ leads to the minimum of the objective function, $I^2$. When the initial guess $\mu_1$ is on the left side of the $\mu^*$, the response is in opposite phase with respect to the perturbation signal. For this case, after integrating $\sigma_1$ (which is the product of perturbation $Asin({\omega}t)$ and response), the obtained integrated value is negative (see Fig.~ \ref{fig:ESschemeConcept}). The integrated value multiplies with negative constant $K$ to obtain a positive value that drives the next estimate of the decision variable, $\mu$, closer to its optimal value, $\mu^*$. Note that the integration will be close to zero when $\mu$ is close enough to the $\mu^*$. If the initial guess is on the other side, e.g. $\mu_2$, the similar estimation process will drive the next estimate closer to the optimal value $\mu^*$ (see Fig.~ \ref{fig:ESschemeConcept}).

\subsection{Droop-based VVC Scheme} 
The droop-based VVC uses a linear mapping from nodal voltages, $V_i$ and respective nodes reactive power generation, $Q_{i,dp}$, see (\ref{eqQdp}). The reactive power dispatch is limited by the available capacity of the smart inverter in (\ref{eqQmax}).  

\vspace{-0.3cm}
\begin{align*}
\small
      Q_{dp}(t+1) = f(& V_t, V_{min}, V_{max}, V_{ref}(t), DB, ... \\
             &... Q_{min}(t),  Q_{max}(t), Q_0(k_n)) 
\end{align*} 
\begin{equation}\label{eqQdp}
\small
f(\bullet) = 
\begin{cases}
   \: \: Q_{max}   &\text{if $ V_t \leq V_{min} $ } \\
   -\dfrac{Q_{max}-Q_0}{V_{l}-V_{min}}(V_t-V_{l})+Q_0   &\text{if $V_{min} \leq V_t \leq V_{l}$ } \\
    \: \: Q_0  &\text{if $V_{l} \leq  V_t  \leq V_{r}$} \\
    -\dfrac{Q_{min}-Q_0}{V_{r}-V_{max}}(V_t - V_{r})+Q_0 & \text{if $ V_{r} \leq V_t \leq V_{max}$} \\
    \: \: Q_{min}  &\text{if $ V_t \geq V_{max} $ } \\
\end{cases}
\end{equation}
where, $DB$ stands for dead band. The left side of bead band is $V_{l}= V_{ref}-DB/2$ and the right side $V_{r}= V_{ref}+DB/2$.
\begin{equation} \label{eqQmax}
\small
 Q_{max}(t) =  -Q_{min}(t) = \sqrt{(S_{PV^{rated}})^2 - (P_{PV}(t))^2} 
\end{equation}

The Volt-VAR droop control is used to maintain the node voltages within a permissible voltage range. Conventionally, the change in voltage is sensed by the local controller. If the node voltage is beyond a permissible value, local controller calculates the required reactive power $Q_{dp}$ from smart inverter using a fixed Q-V slope. The reactive power support will try to maintain the voltages within the limits.  

\section{Extremum-seeking Adaptive-Droop Volt-VAR Optimization (VVO) }

The problem objective is to minimize the total power loss for an unbalanced power distribution system while maintaining the nodal voltages within ANSI \cite{ANSI} limits using only local control of the reactive power dispatch from the smart inverters. 
The notional problem formulation is stated in (\ref{Objective}). 


\begin{equation}
\small
\label{Objective}
\underset{V_{ref}}{\text{minimize:}} \hspace{0.2 cm} \sum_{{ij} \in \mathcal{E}}I_{ij}^2(t) r_{ij}  + \sum_{i \in \mathcal{N}} f_{Vpen}(V_i(t))   
\end{equation}
where,

\begin{small}
\begin{flalign}\label{fVpen}
\small
\nonumber   f_{Vpen}(V_i(t))=
  \begin{cases}
    (0.95-V_i(t)) \times K_{p} &\text{ $ V_i(t) \leq 0.95 $ } \\
    \ \ \ 0                  &\text{ $0.95 < V_i(t) < 1.05$ } \\
    (V_i(t)-1.05)\times K_{p} &\text{ $V_i(t) \geq 1.05$} \\
 \end{cases}
\end{flalign}
\end{small}
\vspace{-0.1cm}

\noindent where, $\mathcal{N}$ and $\mathcal{E}$ represent the set of nodes and lines/edges in a distribution system, respectively; $V_i(t)$ is the voltage at node $i$ at time $t$; $I_{ij}(t)$ is the current flow in line $(ij)$; $r_{ij}$ is the resistance of the branch $(ij)$; $K_p$ is the weighing factor for the voltage violation penalty function, $f_{Vpen}(V_i(t))$. $K_p$ should be of the same order as $I^2$ to weight voltage penalty. 

The problem objective can be decomposed on nodal basis that require only local measurements for decision-making, i.e. $I_{ij}$ and $V_i$ measurements at node $i$. However, since power flow of the network couples the decision variables, the traditional optimal power flow methods require the measurement and model information for the rest of the network to solve the optimization problem detailed in (\ref{Objective}). Note that existing network-level optimization methods including centralized, distributed methods require global (model and measurement) or at  least the measurements from the neighboring nodes to solve the optimization problem. On the contrary, we propose a model-free local control approach to solve the optimization problem in (\ref{Objective}) using only local measurements.


\subsection{Proposed Volt-VAR Optimization Framework}
The proposed approach is a combination of 1) an extremum seeking approach to achieve the global (network-level) objective of minimizing the feeder losses using local perturbation-based controller, and 2) an adaptive droop control approach employed to ensure a stable operation and to regulate fast-changing phenomena such as DER generation variability. The overall framework is detailed in Fig. \ref{fig:proposedvvo}. 

\begin{figure}[t]
\centering
\includegraphics[width=0.99\columnwidth, trim={0 0 0 0},clip]
{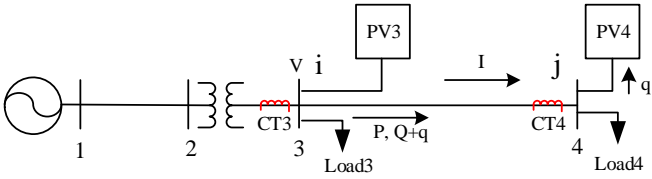}
\vspace{-0.3cm}
	\caption{IEEE 4-bus system with PVs}
\vspace{-0.2cm}
	\label{fig: 4-node}
	\vspace{-0.7cm}
\end{figure}

\subsubsection{Measurements}
The proposed localized extremum seeking controller requires the following measurements: (1) local bus voltage magnitude $V$, (2) branch current magnitude  $I$ from the upstream branch connected to the PV inverter. For example, for the IEEE 4-bus system shown in Fig.~\ref{fig: 4-node}, the required measurements are $V_3$ at node 3 and $I_3$ from CT3 for PV3, and $V_4$ at node 4 and $I_4$  from CT4 for PV4. If the local line impedance matrix is available, we can calculate $S_{loss}$ in (\ref{eq3phaseloss}) as the value of the problem objective to be used in Algorithm 1. If not, the algorithm will take the square of the current magnitude as the problem objective. Note that local active power generation measurement from the PV, $P_{PV}$ is required to calculate the available reactive power capacity, $Q_{max}$ in (\ref{eqQmax}).

\subsubsection{Approach}
The objective of the extremum-seeking controller is to continuously monitor the branch current flow, $I_{ij}$, and search for the gradient of the resulting branch power loss, that is proportional to $I_{ij}^2$. The decision variable for extremum-seeking controller at the node $i$ is $V_{i,ref}$. Note that $V_{i,ref}$ is the optimal voltage that should be maintained at the node $i$ to minimize the power losses. To manage the fast-changing phenomena due to DERs, we propose an adaptive  droop-based Volt-VAR control (VVC) that maintains the reference voltage, $V_{i,ref}$, specified by the extremum-seeking controller in the presence of rapidly varying DER generation patterns. Note that $V_{i,ref}$ identified by the extremum-seeking controller is provided as the input to the droop-controller. To induce a stable response, we also add a steady-state adaptation along with reactive power inertia as detailed in the following sections. 

The proposed local control algorithm is detailed in Algorithm \ref{alg:ESVVO1}. The proposed controller is locally situated at the controllable node, i.e. nodes with smart-inverters. The parameters, $V_{i,ref}$ and $Q_{i,0}$ for the adaptive droop are updated based on the input from the extremum-seeking controller and static state error adaption modules, respectively. The overall approach is divided into two time-scales: the inner loop that executes every 30-sec and the outer-loop that executes every 5-min interval. Essentially, the inner loop is instantaneous, while the outer loop is delayed and operates on aggregated error signal. Note that the extremum-seeking, adaptive VVC, and reactive power hysteresis lie in the inner-loop while the steady state error adaption lies in the outer loop and processes aggregated signal over every $T=10$ time steps. In what follows, we explain the algorithms for the different modules in the proposed extremum-seeking adaptive-droop control approach. 

\begin{algorithm}[!b]
\small
\caption{\small Localized VVO algorithm}\label{alg:ESVVO1}
\SetAlgoLined
    \SetKwInOut{Input}{input}
    \SetKwInOut{Output}{output}
\SetKwInOut{init}{initialize}
\Input{$V_t$, $|I|^2$ or $S_{loss}$}
\Output{$V_{ref}(t)$}
\init{$Q_0(0)$, $V_{ref}(t_0)$, $V_{min}$, $V_{max}$} 
 \While{t in range ($t_0$, $t_{end}$)}{   
  Update $V_{ref}(t)$\;
  \eIf{t = hT, h=1,2,3 ...}{
   Calculate $SSE(k_n)$ from eq(\ref{eqSSE}); \\
   Calculate $Q_0(k_n)$, with limit in [$Q_{min}, Q_{max}$];\\
   Calculate $Q_{dp}(t) = f(V_t, V_{ref}(t), ...)$ from eq(\ref{eqQdp});\\
   Calculate $Q_{PV}(t)$ from eq(\ref{eqQt});\\
   $k_n=k_n+1$;\\
   \eIf{$|Q_0(k_n) - Q_0(k_n-1)| \geq Q_{0Large}$}{
   Reduce outer loop $T$ to $T_{small}$\;
   }{
   Keep same $T$\;
  }
  }
  {
    Calculate $Q_{dp}(t) = f(V_t, V_{ref}(t), ..)$  from eq(\ref{eqQdp});\\
    Calculate PV output $Q_{PV}(t)$ from eq(\ref{eqQt});
  }
  $V_{ref}(t)$ =  $ES_{func}(|I|^2 \ or \ S_{loss} + f_{Vpen}$) \\
 $ES_{func}$ refers to eq(\ref{eqES1}, \ref{eqES2}, \ref{eqES3}) and discrete time sample in \cite{brunton2019data} and search range of $V_{ref}(t)$ is [0.95, 1.05]
 }
\end{algorithm}

\vspace{-0.3cm}


\subsection{Extremum-seeking Control for Network-level Optimization}
The extremum seeking (ES) method is implemented to track optimal solution with continuous small magnitude perturbation in voltage set-points, $V_{i,ref}$, in the droop-controller when generations and loads are changing slowly. Here, the frequency of excitation signal and parameters of washer filters are identical for all smart inverters in the system. The proposed method is using local objective function, thus, there is no significant cross impact from same frequency perturbations of other DERs as observed for different frequency settings in \cite{7350258}.

The problem objective for the controller at node $i$ is obtained upon nodal decomposition of (\ref{Objective}). 
\begin{equation}
\small
\label{Objective_dist}
\underset{V_{i,ref}}{\text{minimize:}} \hspace{0.2 cm} I_{ij}^2(t) + f_{Vpen}(V_i(t))
\end{equation}
\noindent where, $V_{i,ref}$ is the decision variable. It is also the input to the adaptive droop-controller. $I_{ij}$ and $V_i$ are local current and voltage measurements used by the extremum-seeking controller to update the decision-variable, $V_{i,ref}$.  

The line currents and thereby power losses are influenced by reactive power injection $q_i$ at the individual node. The adaptive droop defines the relationship between node voltage, $V_i$ and the reactive power dispatch, $q_i$. The extremum-seeking identifies the $V_{i,ref}$ for the adaptive droop-controller at each time-step, so that the corresponding reactive power dispatch, $q_i$, minimizes the local line losses for the branch $\{ij\}$ while maintaining the nodal voltage within the respective upper and lower bounds. At every time step, $T$, the ES method adds a continuous perturbation to $V_{i,ref}$, to perturb the local reactive power dispatch, $q_i$, to minimize the problem objective in (\ref{Objective_dist}). The gradient estimation of ES will drive $V_{i,ref}$ to  optimize for (\ref{Objective_dist}) using the perturbation and response from the system as detailed in Section II.A (see Fig. \ref{fig:ESschemeConcept}).

\vspace{-0.4cm}
\subsection{Adaptive Droop Control for Fast-changing Phenomenon}
\begin{figure}[t]
\centering
\vspace{-0.3cm}
\includegraphics[width=.99\columnwidth, trim={0cm 0cm 0cm 0cm},clip]{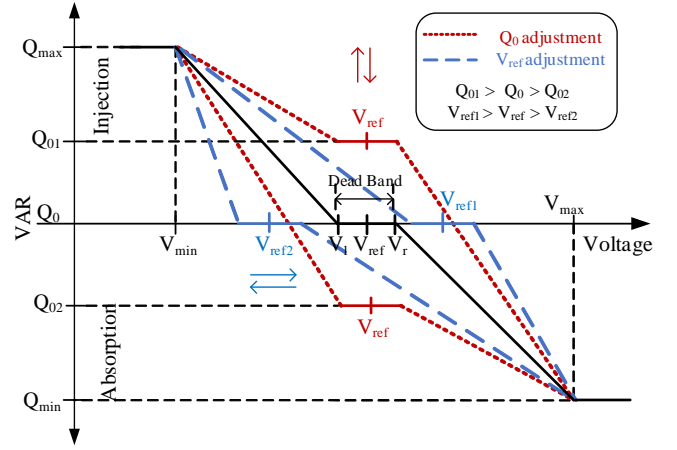}
\vspace{-0.3cm}
	\caption{$V_{ref}$ and $Q_0$ adjustment in Volt-VAR control}
	\label{fig:VrefQ0Adjust} 
	\vspace{-0.8cm}
\end{figure}

Compared to the conventional VVC, an adaptive VVC allows for the parameters, $Q_0$ and $V_{ref}$ in (\ref{eqQdp}) to be adjustable (see red and blue lines in Fig.~\ref{fig:VrefQ0Adjust}). Thus, an adaptive droop VVC allows for network-level optimization by generating the desired reactive power to maintain optimal voltage setpoints as identified by the ES controller. First we check the parameters, $V_{min}, V_{max}, V_{ref}, Q_{min}, Q_{max}, Q_0$, which are required to determine a droop curve as shown in Fig. \ref{fig:VrefQ0Adjust}. The $Q_{min}, Q_{max}$ are determined by the kVA rating of the smart inverter  and active power output of a DER at a given time from (\ref{eqQmax}). The value for $V_{min}, V_{max}$ are the adjustable parameters which are specified in IEEE 1547 standard \cite{8332112}. 
In the proposed approach, $V_{ref}$ is determined and periodically updated by a local extremum seeking optimization method to achieve network-level optimization. Another problem with the droop-controllers is that they introduce reactive power oscillations due to sharp slope and steady state error(SSE) \cite{singhal2018real}. Thus, here we introduce reactive power inertia through low pass filter to mitigate sudden large $Q_{dp}$ variance as shown in Fig.\ref{fig:proposedvvo}, and SSE adaption to reduce the setting errors highlighted in red lines shown in Fig.\ref{fig:VrefQ0Adjust}.   

\subsubsection{Reactive Power Hysteresis}
The low-pass filter mitigates sudden reactive power changes and oscillation caused by droop control, thus this filter is added after the Volt-VAR droop control scheme \cite{jahangiri2013distributed}. The final reactive power generation, $Q_{dp}$, is determined by the previous time step reactive power dispatch and the results from the VVC function in (\ref{eqQdp}).  
The settings of $V_{ref}, Q_0, Q_{min}, Q_{max} $ are time varying variables, which is different from traditional Volt-VAR control. 
\begin{equation}\label{eqQt}
\small
    Q_{PV}(t+1) = (1-\mu) Q_{PV}(t) + \mu Q_{dp}(t+1)
\end{equation}

\subsubsection{Steady State Error}
The SSE is defined as the difference between $V_{ref}$ and steady state voltage in (\ref{eqSSE}) \cite{singhal2018real}. By measuring the real-time SSE, we can obtain an  adaptive $Q_0$ which helps in mitigating the reactive power oscillations. The oscillation is caused by significant difference between the droop control setting, $Q_0$, and the required reactive power, $Q_{dp}$, to maintain the nodal voltage close to $V_{ref}$. By default, $Q_0$ is zero as per IEEE-1547 standard. However, for load node, the $Q_{dp} > 0$ is needed to raise the nodal voltage to $V_{ref}$.  Essentially, the difference in $Q_{dp}$ and $Q_{demand}$ leads to oscillations in the droop-controller. The SSE reduces the gap by adaptively setting the $Q_0$ and adjusting the slope to mitigate oscillations, see Fig. \ref{fig:VrefQ0Adjust}. 

\begin{small}
\begin{eqnarray}\label{eqSSE}
\centering
SSE(k_n ) &=& \sum_{t=k_nT+1}^{(k_n+1)T} (V_{t}-V_{ref,t})   \\
Q_0(k_n) &=& Q_0(k_n-1) -K_Q *SSE(k_n), k_n=1,2 ... 
\end{eqnarray}
\end{small}
\noindent Where, the averaging time-interval, $T = 10$ time-steps, and $K_Q$ is positive number in the same order as the inverter capacity when per-unit values of voltages are used. Note that when the required change in $Q_0$ is larger than a pre-specified threshold, say $Q_{Large}$, we shorten the averaging time intervals $T$, for instance from 10 to 8, as shown in Algorithm \ref{alg:ESVVO1}. This helps track fast load changes.

\section{Convexity of Power Loss Function}
In this section, we check the convexity conditions for the objective functions to ensure the proposed algorithm will converge to global minimum. We introduce the mathematical formulation for a single and three-phase power loss based on branch flow model. Next, valid approximations are proposed to reduce the original formulation into a linear voltage penalty function and an equivalent quadratic formulation to prove the convexity of the objective function. Here, to simplify the problem, we assume $V_i \approx V_j = V$. Also, due to the linear droop mapping between $V$ at node $j$ and reactive power generation $q$ as shown in  Fig. \ref{fig:VrefQ0Adjust}, the voltage penalty function is convex w.r.t. $q$. Thus, it does not change the convexity of the combined objective function. 
\vspace{-0.3cm}
\subsection{Objective Function Convexity in Single Phase System}
For a single phase system, the objective of loss minimization is simply represented as the $I^2$ minimization. In order to prove convexity of  the function, we need to determine the conditions for which the second derivative of $I^2$ is positive i.e. $\pdv[2]{I^2}{q}\geq 0$ w.r.t to control variable, here,  reactive power generation $q$ from photovoltaic (PV). If the function is convex, one can claim that the extremum seeking helps in achieving the optimal solution. The notations, $V, P, Q+q$ are the voltage, active power, reactive power measurements at node $i$ of branch $\{i,j\}$ respectively. The impedance of the branch is given as $z = r+\iota x, \text{where, }\iota=\sqrt{-1} $. The current of branch $\{ij\}$ is noted as $I$ and $q$ is reactive power absorbed by the PV at a node $j$ (see Fig.~\ref{fig: 4-node}). Take PV4 at node 4 for instance, the branch current is measured by CT4 and voltage at node 4 are utilized for PV4 to make the  decision on $q$ dispatch. Similarly for PV3, the branch current measurement from CT3 and voltage measurement at node 3 are inputs to the proposed ES algorithm for $q$ dispatch. Due to linear mapping between $q$ and $V$ via droop control, the convex relationship for problem objective and control variable, $I^2$ and $V$, holds.   

The relationship between the current, active and reactive power flow and the node voltage is given as:
\begin{equation} \label{eqI}
\small
     I^2V^2=P^2+(Q+q)^2
\end{equation}
Taking $y = V^2$ the second derivative of $P$ and $Q+q$ w.r.t. $q$ we obtain,
\begin{small}
\begin{eqnarray}
 \pdv[2]{(P^2/y)}{q} =\dfrac{2 P}{y} \pdv[2]{P}{q} -  \dfrac{P^2}{y^2} \pdv[2]{y}{q}  + \dfrac{2}{y} M  \label{eqp}  \\
\pdv[2]{((Q+q)^2/y)}{q} = -\dfrac{(Q+q)^2}{y^2}\pdv[2]{y}{q}+\dfrac{2(Q+q)}{y} \pdv[2]{Q}{q} + \dfrac{2}{y}N \label{eqq}
\end{eqnarray} 
\end{small}
\noindent where,
\begin{small}
\begin{flalign}\label{eqdel12}
\noindent
\nonumber M = \left(\pdv{P}{q} - \dfrac{P}{y}\pdv{y}{q}\right)^2 \hspace{0.05cm} \text{and} \hspace{0.05cm} 
N = \left[ \left(\pdv{Q}{q}+1\right) - \dfrac{(Q+q)}{y}\pdv{y}{q}\right]^2 
\end{flalign}
\end{small}
Using (\ref{eqI}),  (\ref{eqp}) and (\ref{eqq}) the partial derivative of $I^2$ w.r.t. $q$ is given as:
\vspace{-0.1cm}
\begin{equation}\label{eqIq}  
\small
\begin{aligned}
 \pdv[2]{I^2}{q} &= \dfrac{2 P}{y} \pdv[2]{P}{q} -  \dfrac{P^2}{y^2} \pdv[2]{y}{q} -\dfrac{(Q+q)^2}{y^2} \pdv[2]{y}{q} \\
   & +  \dfrac{2(Q+q)}{y} \pdv[2]{Q}{q} + \dfrac{2}{y} M + \dfrac{2}{y}N 
  \end{aligned}
\end{equation} 
Next, by assuming $\pdv[2]{y}{q} \leq 0, \pdv[2]{P}{q} \geq 0, \pdv[2]{Q}{q} \geq 0$ and $\small{P = P_{L}+I^2r,  Q = Q_L + I^2x}$ we obtain, 


\begin{equation}\label{eqIq2}
\small
\begin{aligned}
 \pdv[2]{I^2}{q} &= r\dfrac{2 P}{y} \pdv[2]{I^2}{q} + x\dfrac{2(Q+q)}{y} \pdv[2]{I^2}{q} - \dfrac{P^2}{y^2} \pdv[2]{y}{q} \\
 & -\dfrac{(Q+q)^2}{y^2} \pdv[2]{y}{q} + \dfrac{2}{y} M + \dfrac{2}{y} N  
  \end{aligned}
\end{equation}


Taking the common terms together in (\ref{eqIq2}), we simplify the term as in (\ref{eqIfinal}). The convexity of  (\ref{eqIfinal}) can be assured if $({1-2rP/y-2x(Q+q)/y} ) \geq 0$.
\begin{equation}\label{eqIfinal}
\small
\pdv[2]{I^2}{q} = \dfrac{K}{1-2rP/y-2x(Q+q)/y}
\end{equation}
\begin{equation}\label{eqIfinalK}
 \small
 K= - \dfrac{P^2}{y^2} \pdv[2]{y}{q} -\dfrac{(Q+q)^2}{y^2} \pdv[2]{y}{q} + \dfrac{2}{y} M + \dfrac{2}{y} N   \geq 0 
\end{equation}

Next, as the distribution system is a three-phase unbalanced system, we derive the condition for which the power loss term can hold convexity property.

\vspace{-0.1cm}
\subsection{Three-phase Unbalanced and Balanced System}
In a three-phase unbalanced system ($\Phi \in \{a,b,c\}$), total  loss in a branch is sum of loss due to self and mutual impedance. Mathematically, the total power loss in a three-phase line $(i,j)$ is given by (\ref{eq3phaseloss}), where $m,n$ are phases of node i and j respectively $\{(m,n): m\in \Phi_i, n\in \Phi_j\}$. $z_{mm}$ and $z_{mn}$ are the self and mutual impedance.  Note, for simplification, the line reference $(i,j)$ is removed from the equations. In (\ref{eq3phaseloss}), the power loss consists of six mutual loss terms and three self loss terms. The loss term due to mutual current components are described in (\ref{eqIpq}).
\begin{equation}\label{eq3phaseloss}
\small
  S_{loss} = \sum_{m,n \in \Phi} z_{mn}I_{mn}I_{mn}^*  
\end{equation}

Let for a line $(i,j)$ phase A is reference phase with apparent power $S_a$ equal to $P+\iota (Q+q)$, and the other two phases B and C power flow are  given as $ P+\Delta P_1 , P+\Delta P_2$ and $Q+\Delta Q_1 + q , Q+\Delta Q_2 +q$ respectively. Assuming, phase angle difference among all the phases are $120^0$, the relationship among the phase current and power flow in a line $(i,j)$ $ \forall m,n \in (a,b,c)$ is given as:


\begin{small}
\begin{flalign}\label{eqIpq}
I_{m}^{*}I_{n} + I_{n}^{*}I_{m} =  \left(\frac{S_m}{V_m} \right)^*\left(\frac{S_n}{V_n} \right) + \left(\frac{S_n}{V_n} \right)^*\left(\frac{S_m}{V_m} \right) 
\end{flalign}
\end{small}


The total power loss in a line w.r.t.  power flow can be obtained using (\ref{eq3phaseloss}) and (\ref{eqIpq}) as shown in (\ref{eqS3ph}).
\begin{small}
\begin{flalign} \label{eqS3ph}
\small
\nonumber S_{loss} =& \{[P^{2}+(Q+q)^2]( z_{aa}+ z_{bb}+ z_{cc}- z_{ab}- z_{ac}- z_{bc})\\
\nonumber    & +[P \Delta P_{1}+(Q+q)\Delta Q_{1}](2z_{bb}- z_{ab}- z_{bc}) \\
\nonumber     & + [P \Delta P_{2}+(Q+q)\Delta Q_{2}](2z_{cc}- z_{ac}- z_{bc}) \\
 \nonumber    & + \sqrt{3}[P\Delta Q_{1} -(Q+q)\Delta P_{1}](z_{ab}- z_{bc}) \\
  & + \sqrt{3}[P\Delta Q_{2} -(Q+q)\Delta P_{2}](z_{bc}- z_{ac})\} /y
\end{flalign}
\end{small}

The second order partial derivatives $\small{\pdv[2]{(P\Delta P/y)}{q}\geq 0}$, $\small{\pdv[2]{((Q+q)\Delta P/y)}{q}\geq 0}$, $\small{\pdv[2]{(P\Delta Q/y)}{q}\geq 0}$, $\small{\pdv[2]{((Q+q)\Delta Q/y)}{q}\geq 0}$ based on following assumptions
$\small{\Delta P_1, \Delta Q_1, \Delta P_2, \Delta Q_2 \geq 0}$, $\small{\pdv[2]{P}{q}, \pdv[2]{Q}{q}\geq 0}$, $\small{\pdv[2]{y}{q} \leq 0}$, $\small{\pdv[2]{\Delta P_1}{q},\pdv[2]{\Delta Q_1}{q}, \pdv[2]{\Delta P_2}{q},\pdv[2]{\Delta Q_2}{q} \approx 0}$. These conditions $\small{\pdv[2]{P}{q}, \pdv[2]{Q}{q}\geq 0}$, $\small{\pdv[2]{y}{q} \leq 0}$ are provided and validated in \cite{7350258}.

For the three-phase balanced system, we assume all three phase power flow are $P_{a}=P_{b}=P_{c}=P$ and $Q_{a}=Q_{b}=Q_{c}=Q+q$. Thus the condition for convex function of the loss term is same as that of a single phase network.


\section{Results and Discussions}
The proposed VVO approach is validated using  two test feeders: IEEE 4-bus and IEEE 123-bus systems \cite{TestFeeder}. The IEEE 4-bus system is assumed to be a three-phase balanced system. Here, first we investigate how the proposed method works to maintain the voltages within a permissible range given by ANSI \cite{ANSI} and mitigate control oscillation by comparing with original droop based VVC. Next, we observe the interactive responses of two smart inverters to observe if there are any conflicts between the control signals generated by the smart inverters with the extremum seeking control. For the IEEE 123-bus unbalanced system, we demonstrate that the proposed controller is able to achieve the desired objective of network-level loss minimization using only local control methods. The results obtained from the proposed approach are compared against the optimal solutions obtained using a centralized VVO approach for unbalanced three-phase system from our prior work \cite{bileveljha}. It is shown that the proposed local control method closely follows the optimum solutions obtained from the centralized VVO. 

\begin{figure}[t]
\centering
\includegraphics[width=0.9\columnwidth, trim={0 0 0 0},clip]{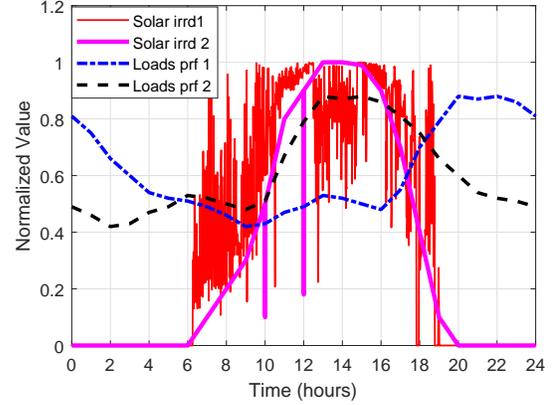}
\vspace{-0.1cm}
	\caption{Normalized solar irradiation and loads in 30s interval. (Solar1: high variance, Solar2: smooth profile with spikes at 10h and 12h)}
\vspace{-0.3cm}
	\label{fig:SolarLoadProfile}
\end{figure}

\begin{figure}[t]
\centering
 \includegraphics[width=0.5\textwidth]{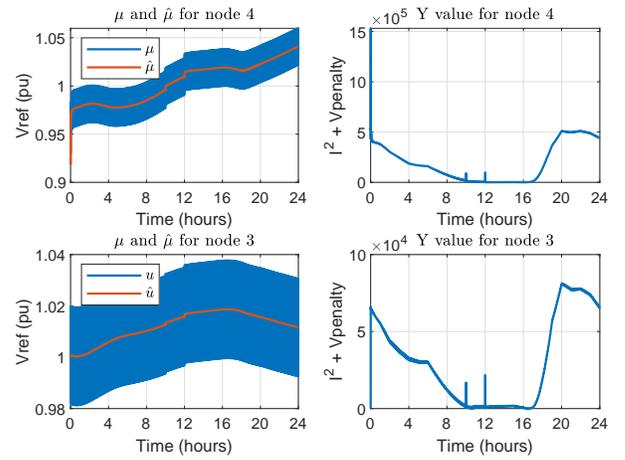}
\caption{$V_{ref}$ and objective y tracked by proposed VVO }
	\label{fig:Vref4}
	\vspace{-1cm}
\end{figure}

\begin{table}[t]
		\centering
		\caption{PVs and loads size settings}
		\label{tab:4nodesPVloads}
\vspace{-0.2cm}
		\label{singletable}
		\begin{tabular}{c|c|c}
			\toprule[0.4 mm]
			\hline
			{Load/PV} & $Rated P+Q$ & $Rated S$ \\
			\hline
			\hline
			{\hspace{-4pt}Load 3}&  1400 kW + 678 kVar (Pf 0.90) & 1556 kVA \\
			  \hline
            {\hspace{-4pt}Load 4}& 5400 kW  +2615 kVar (Pf 0.9) & 6000 kVA \\
            \hline
            {\hspace{-4pt}Load 4*Rev}& 5400 kW - 1775 kVar (Pf -0.95) & 5684 kVA \\
            \hline
             {\hspace{-4pt}PV 3}& 2000 kW  & 2400 kVA \\
            \hline
            {\hspace{-4pt}PV 4}& 3000 kW  & 3600 kVA  \\
            			\toprule[0.4 mm]
			\end{tabular}
		\vspace{-1.8cm}
\end{table}
\subsection{Proposed Volt-VAR Control in IEEE 4-bus System }
The IEEE 4-bus system is modified to test the proposed method by adding a load at node 3 and PVs at node 3 and 4 (see Fig. \ref{fig: 4-node}). The parameters are shown in Table \ref{tab:4nodesPVloads}. The smooth solar profile Solar2 with two spikes at 10th and 12th hours for short-time cloud covers, load profiles 1 and 2 are modified from OpenDSS PV test example shown in Fig. \ref{fig:SolarLoadProfile} \cite{OpenDSS}. The smart inverters are sized to be 1.2 times of the rated active power. 

\begin{figure}[t]
\centering
\includegraphics[width=0.5\textwidth]{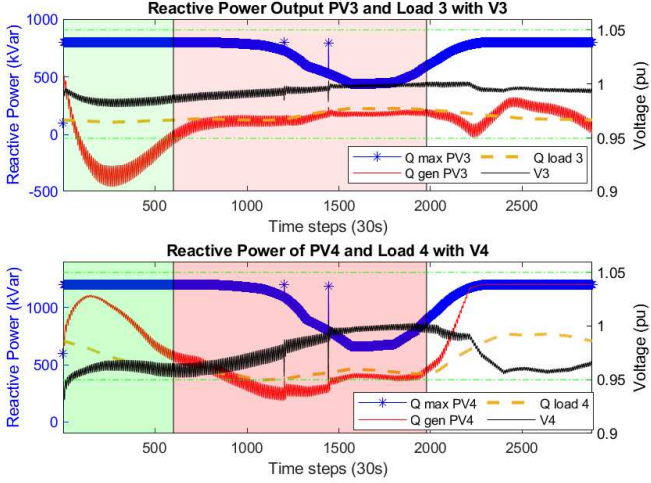}
\vspace{-0.5cm}
	\caption{Reactive powers of PVs and voltages at node 3 and 4}
	\label{fig:Q3n4}
	\vspace{-0.7cm}
\end{figure}

\subsubsection{Voltage violation penalty and power loss minimization } 
The change in $V_{ref}$ and its corresponding objective function value  with time is shown in Fig.~\ref{fig:Vref4}. Here, $\mu$ is $V_{ref}$ and $\hat{\mu}$ is $V_{ref}$ estimation, which corresponds to the variables in (1)-(3). The convexity condition in (\ref{eqIfinal}) is checked and satisfied at node 4 every time step, which assure the proposed method will drive the local power loss as low as possible. We observe the objective function value is high at initial points due to voltage violation (0.87 pu) at node 4 and drop immediately upon adjusting the $V_{ref}$. It can be observed from Fig. \ref{fig:Q3n4} and \ref{fig:volt3n4} that the initial voltage violations are mitigated using the reactive power dispatch from PVs at nodes 3 and 4.  

The reactive power responses for smart inverters at nodes 3 and 4 for the 24-hour simulation is shown in Fig. \ref{fig:Q3n4}. We identify three time sections to highlight the reactive power control. During the initial time-period, shown in light green color background, the proposed method mitigates the initial low voltage issue at node 4 by ramping up the reactive power dispatch from the PV installed at node 4, then gradually tracking the local reactive power load demand. During the next interval shown in light red, when the nodal voltages are within the limits, the controller tracks the reactive power load demand to minimize the local branch currents $I^2$. Further, we also demonstrate its ability to manage the cloud transients at $10^{th}$ and $12^{th}$ hour using  Solar irrd 2 profile in Fig. \ref{fig:SolarLoadProfile}. The third part of the plot shown in white color demonstrates the ability of the proposed controller to deal with fast ramping loads by utilizing all reactive power reserve to maintain voltage and reduce loss. 

We also test the approach with other load models. In IEEE 4-bus test system, we change loads 3 and 4 from constant PQ model to 50\% constant current and 50\% constant impedance load model for both P and Q components. It was observed that the proposed approach is capable of tracking the reactive power demand and reducing the power loss, which is similar to results as shown in Fig.~\ref{fig:Q3n4}.

Note that the time interval for the operation of inner loop control is selected to be 30s for the following two considerations. First, we assume the sampling rate for the SCADA system is 4-10s \cite{8039196,8784481}. Second, we assume that the controller will take an additional 10-20s for computation and for modifying the voltage setting accordingly. For the outer control loop, we choose a time interval that is 10 times of the inner loop to allow for the tracking of slow changes. The inner-loop can be made faster for example running at every 10s time interval.

\subsubsection{Perturbation frequency selection} We apply same perturbation frequency to all PV inverters.  Reference \cite{7350258} uses feeder head active power measurement ($P_0$), that is a composition of the frequency perturbations at individual inverters. Thus, each inverter needs to filter out the frequency perturbations of other inverters (in the $P_0$ measurement) to estimate the gradient. Regarding the proposed method, the problem objective for each ES controller is the square of local branch current magnitude, $|I|^2$. Thus, all inverters see a different objective function unlike a common feeder head active power measurement ($P_0$) used in \cite{7350258}. 

\begin{figure}[t]
\centering
\includegraphics[width=0.98\columnwidth, height=2.15in, trim={0.0cm .0cm 0.0cm
0.0cm},clip]{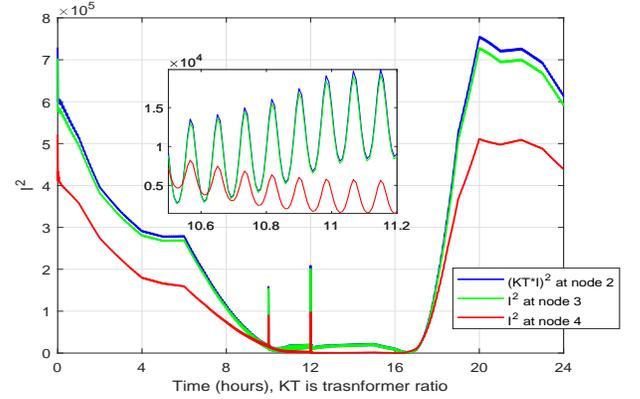}
	\caption{Perturbations impact on $|I|^2$ at node 2, 3 and 4}
	\label{fig:Perturbance impact}
	\vspace{-0.50cm}
\end{figure}


For our scenario, the perturbations in local $V_{ref}$ cause perturbations in reactive power measurements, $q_p$. In this case, the local and neighboring reactive power perturbations cause the combined perturbations in the reactive power at the inverter location ($q_{cmb}$). The combined reactive power perturbation results in perturbations in local current (the problem objective) at the inverter location. Take IEEE 4-bus system as an example;  (\ref{eq:Vref3}-\ref{eq:QcmbtoI}) show how the perturbations at node 3 from $V_{ref3}$ are transferred to local reactive power perturbations ($q_{3p}$), that are then combined with $q_{34p}$ from node 4 to obtain the net reactive power perturbation ($q_{cmb}$). The net reactive power perturbation ($q_{cmb}$) then affects the perturbations for the branch current measurements, $I_{3p}$ as per (25).

\vspace{-0.2cm}
\begin{small}
\begin{flalign}
\centering
\label{eq:Vref3}
&V_{ref3}=\hat{\mu_{3}}+A_3\sin(\omega t  +\theta_{3})\\ \label{eq:Q3p}
&V_{ref3} \xrightarrow{Volt-VAR} q_{3p}\sin(\omega t  +\theta_{3})\\ \label{eq:Q34P}
&V_{ref4}\xrightarrow[\text{line 3-4}]{Volt-VAR} q_{34p}\sin(\omega t +\theta_{4}) \\ 
\nonumber &q_{cmb}\sin(\omega t +\theta_{cmb})=q_{3p}\sin(\omega t  +\theta_{3}) + q_{34p}\sin(\omega t +\theta_{4}) \ \   \\ \label{eq:Qcmb}
& \text{where, }\theta_{3}=\theta_{4}=\theta_{cmb}=0 \\ 
&q_{cmb}\sin(\omega t +\theta_{cmb})\rightarrow|I_{3p}|^2\sin(\omega t +\theta_{I3})\rightarrow\eta \ in  \ (2)  \label{eq:QcmbtoI}
\end{flalign}
\end{small}
\vspace{-0.4cm}

\noindent where, $q_{3p}$ is the reactive power perturbation (magnitude) from PV3, and $q_{34p}$ is the reactive power perturbation at node 3 due to PV at node 4 (PV4). Next, $q_{cmb}$ is the combined reactive power perturbation observed at node 3 and $I_{3p}$ is the perturbation observed in branch current. Note that $\eta $ in  (\ref{eqES2}) is referred to the extremum seeking equations (see Fig.~\ref{fig:proposedvvo}).

\begin{figure}[t]
\centering
\includegraphics[width=0.98\columnwidth, height=2.15in, trim={0.0cm .0cm 0.0cm 0.0cm},clip]{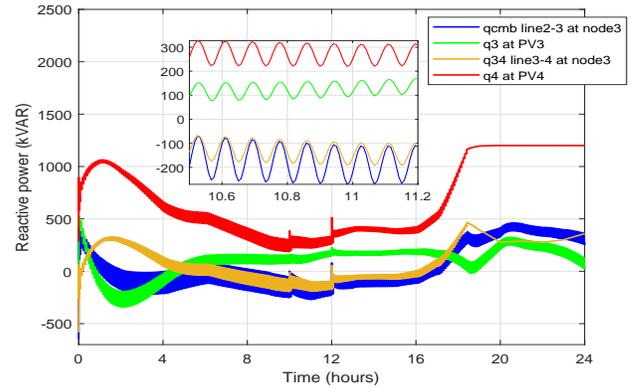}
	\caption{Combined perturbation $q_{cmb}$ from PV3 and PV4}
	\label{fig:Q PerturbanceInPhase}
	\vspace{-1.20cm}
\end{figure}

We simulated a simple case to observe the effects of perturbations in neighboring buses on the local PV bus. From (\ref{eq:Qcmb}), the combined perturbation $q_{cmb}$ has the same phase as the local perturbations, $q_{3p}$, but of different magnitudes. In Fig.~\ref{fig:Q PerturbanceInPhase}, we can see the combined reactive power perturbation, $q_{cmb}$, $q3$ from PV3 and $q4$ from PV4 are of same phase. The only difference between the combined perturbation $q_{cmb}$ at node 3 and perturbation $q3$ from PV3 is in the magnitude. The parameter $K$ can be easily adjusted to this difference (Fig.~\ref{fig:proposedvvo}). Thus, each ES controller can be equipped with perturbation in same frequency. We do not need to assign each ES controller with specific frequency filters to obtain the differential response from each PV as was required \cite{7350258}.


\subsubsection{Control oscillation mitigation}
\begin{figure}[t]
\centering
\includegraphics[width=0.98\columnwidth,trim={0.0cm .0cm 0.02cm 0.0cm},clip]{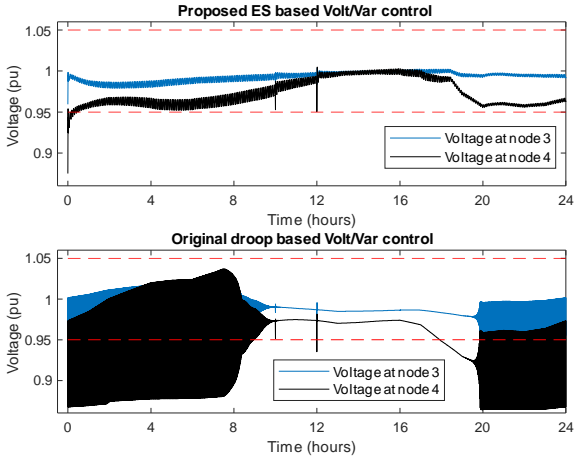}
\vspace{-0.2cm}
	\caption{Voltages comparison between proposed VVO and droop based VVC (oscillation + violation)}
 \vspace{-1.40cm}
	\label{fig:volt3n4}
\end{figure}

To introduce control instability, we narrow down the gap between $V_{min}$ and $ V_{max}$ to [0.88, 1.12] pu in (\ref{eqQdp}); this increases the slope of the droop curve. The proposed VVO is compared against the traditional droop-based VVC with the same $V_{min}$ and $ V_{max}$ parameters. The traditional VVC is set to $Q_0=0, V_{ref}= 1$. The instability in VVC is observed for most of the day and voltage violations are observed between hours 0 and 8 at node 4. Meanwhile, the proposed extremum-seeking adaptive droop approach observes no voltage violation, see Fig. \ref{fig:volt3n4}. 
Note that in the traditional VVC, the oscillations disappear between hours 10 and 19. This is because, $Q_{max}$ is smaller given higher magnitude of $P_{PV}$ in this time interval, which reduces the slope of the droop curve. 

\begin{figure}[t]
\centering
\includegraphics[width=0.99\columnwidth,trim={0cm 0cm 0cm 0cm},clip]{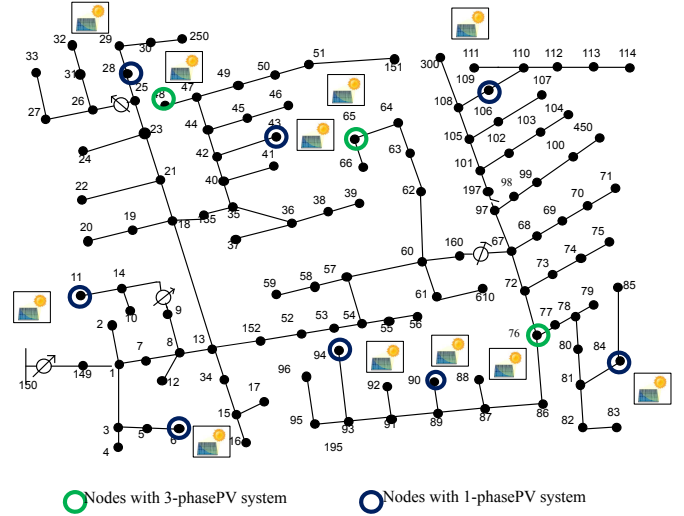}
	\caption{IEEE 123-bus test feeder with PV installations}
	\label{fig:IEEE123}
	\vspace{-1.10cm}
\end{figure}

\subsection{Validation using IEEE 123-bus Unbalanced System}
The IEEE 123-bus feeder is modified by adding multiple PVs to validate the proposed method in a unbalanced three-phase power system as shown in Fig. \ref{fig:IEEE123}. Single phase nodes  are connected with  24 or 48 kVA and  three-phase nodes with  60 or 120 kVA ratings of smart inverters interfaced with PVs system. The solar irradiation 1, load profile 1 are executed for all PVs and loads respectively, while the highly variable solar irradiation 2 profile is implemented to validate the performance of the proposed ES method (see Fig. \ref{fig:SolarLoadProfile}).  For the simulation purpose we have set the substation voltage to 1.05 pu. 
 
The proposed approach is compared against the centralized VVO proposed in \cite{bileveljha} and the traditional droop-based VVC with $V_{ref}=1.0$ pu. Here, we assume that the traditional VVC is equipped with SSE adjustment function to ensure a stable operation as proposed in \cite{singhal2018real}.
The effectiveness of the proposed VVO method is validated by demonstrating its ability to optimize the feeder losses while satisfying the node voltage constraints and reach to the same optimum solution as obtained by the centralized VVO. However, unlike the centralized VVO, the proposed approach is also able to manage the highly variable PV generation scenario and maintain feeder voltage within the acceptable ANSI limits. 

\begin{table}[h]
		\centering
		\caption{Powerloss and Voltage comparison of three-methods}
		\label{tab:ThreeMethods}
\vspace{-0.2cm}
		\label{singletable}
		\begin{tabular}{c|c|c|>{\columncolor[gray]{0.8}}c|c}
			\toprule[0.4 mm]
			\hline
			{Solar Irr} & $Metrics$& $Centralized$ & $ Proposed $ & $V_{ref}=1pu$ \\
			\hline
			\hline
			{Smooth}&{\hspace{-4pt}Powerloss}&  719.09 kWh &    770.94 kWh & 1178.3 kWh \\
			  \hline
            {Smooth}&{\hspace{-4pt} Lowest V}& 0.999 pu & 0.989 pu & 0.967 pu \\
            \hline
            {HighVar}&{\hspace{-4pt}Powerloss}& 716.32 kWh & 760.27 kWh & 1123.90 kWh \\
            \hline
            {HighVar}&{\hspace{-4pt} Lowest V}&  0.999 pu &  0.985 pu  & 0.967 pu \\
           
            			\toprule[0.4 mm]
			\end{tabular}
		\vspace{-0.1cm}
\end{table}



\subsubsection{Power loss comparison and voltage profiles comparison} 
\begin{figure}[t]
\centering
\vspace{-0.3cm}
\includegraphics[width=0.98\columnwidth,height=2.15in]{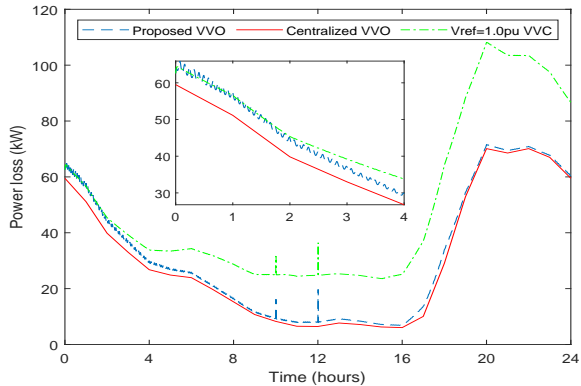}
\vspace{-0.1cm}
	\caption{Power loss comparison among proposed VVO, default $V_{ref}$ VVC and centralized VVO method with solar2 profile }
\vspace{-0.50cm}
	\label{fig:powerloss123}
\end{figure}

The power loss profile of proposed method is close to that of centralized method, while the traditional VVC with default $V_{ref} = 1.0 pu$ causes higher power losses, especially during higher load demand (18-24h) as shown in Fig. \ref{fig:powerloss123}. The little difference between power loss profiles of the proposed VVO and centralized VVO is caused due to the slow-tracking of the extremum seeking method. The difference is however, extremely small (see Table \ref{tab:ThreeMethods}). The centralized VVO coordinates all PVs to generate close to full available reactive power to minimize reactive power demand from feeder head and thus reduce power loss. While, the proposed localized VVO achieves near-optimal loss minimization using only local measurements. Compared to this, using only traditional VVC results in significantly higher power losses thus emphasizing the value of the proposed extremum-seeking adaptive VVC. Assuming the electricity price is \$0.08 per kWh with daily power loss shown in Table III, the proposed method will save 407kWh daily, about \$11,894/year, compared to the default case with $V_{ref} = 1.0 pu$. On the contrary, for a highly variable solar profile, the proposed method will save 363 kWh per day, about \$10,617/year, compared to case of default $V_{ref} = 1.0 pu$. Table.~\ref{tab:ThreeMethods} also shows the lowest node voltages observed for each node. None of the methods report any voltage violations. 

\begin{figure}[t]
\centering
\includegraphics[width=0.98\columnwidth, height=2.15in]{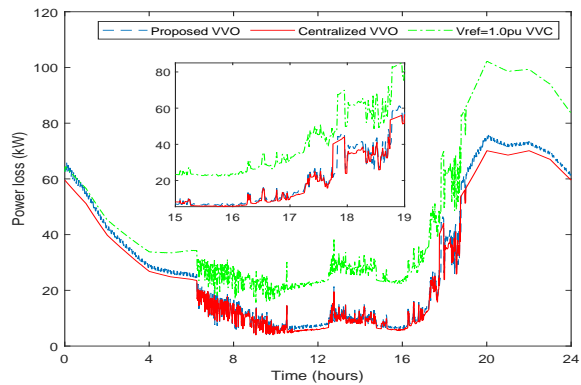}
\vspace{-0.1cm}
	\caption{Power loss comparison between proposed VVO and default $V_{ref}$ VVC with highly variable solar irradiation 1 profile }
\vspace{-0.3cm}
	\label{fig:powerlossHVsolar}
	\vspace{-1.2cm}
\end{figure}

\subsubsection{Highly variable solar irradiation scenario}
Next, we demonstrate the ability of the proposed controller with high PV variability. The original highly variable solar irradiation data (solar irrad 1 in Fig.~\ref{fig:SolarLoadProfile}), is interpolated to 30 seconds time-interval. However, the centralized VVO is solved for every 5-min interval and the $V_{ref}$, is fixed during that 5-min time interval. The power loss profiles of the proposed VVO and centralized VVO for high variability cases in shown in Fig.~\ref{fig:powerlossHVsolar}. As it can be observed, the proposed controller achieves near optimal power loses even for cases with high PV variability without resulting in any voltage violations (see Table III).
\subsubsection{Objectives $S_{loss}$ and $|I|^2$  comparison in highly variable solar irradiation scenario}
It is to be noted that directly minimizing $S_{loss}$ rather than  using $|I|^2$ as the problem objective would result in a more accurate solution. However, the calculation of $S_{loss}$ requires the knowledge of branch impedance. Here, we show via simulation that $|I|^2$ can be used as the problem objective if the branch impedances are not available. Specifically, for the IEEE 123-bus test system, both $S_{loss}$ and $|I|^2$ as the problem objective lead to similar results. For 24-hour simulation, the power loss of using $|I|^2$ as objective is 760.27kWh, while that of using $S_{loss}$ as objective it is 746.19 kWh. We see that using branch impedance matrix will help further reduce power loss, although using branch current, $|I|^2$, provides close enough results.

\subsubsection{Performance with Legacy Voltage Control Devices}
Note that the legacy voltage regulation devices will adjust tap positions and connect/disconnect statuses of the capacitor banks switches in response to their local control settings. The status changes of those devices will change local state variables at the inverter nodes (node voltages and branch currents) that will affect the output of the proposed ES algorithm. Specifically, the PV inverters will adjust $V_{ref}$  accordingly to minimize the current magnitude in order to reduce the branch power loss. Since, the ES approach uses local state variables in the proposed local Volt-VAR algorithm, the proposed controllers will respond to the updated system status and will still be able to achieve the local branch power loss objective. 

\begin{figure}[t]
\centering
\includegraphics[width=0.98\columnwidth, height=2.15in]{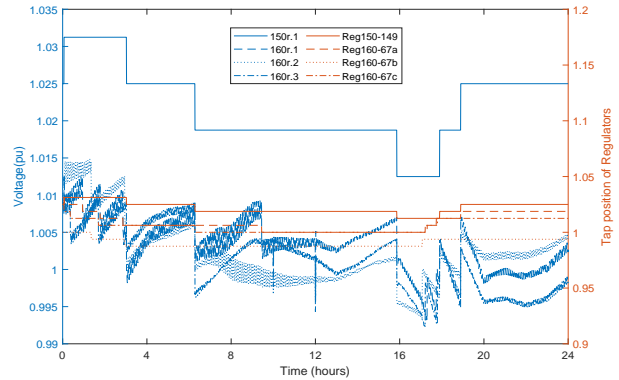}
\vspace{-0.2cm}
	\caption{Voltages and taps changes of regulators}
\vspace{-0.3cm}
	\label{fig: VoltagesTapReg}
	\vspace{-0.2cm}
\end{figure}

We have simulated test case to validate the functionality of the proposed Volt-VAR control scheme in the presence of voltage regulator operations. 
With 1.0 pu feeder head voltage, we observe the regulator operations. The voltage regulator setting is 1.0 pu for all regulators (Reg 150-149, Reg 9-14a, Reg 25-26a, Reg 25-26c, Reg 160-67a, Reg 160-67b, Reg 160-67c). The tap change operations for the duration of a day is shown in Fig. \ref{fig: VoltagesTapReg}.
 
\begin{figure}[t]
\centering
\includegraphics[width=0.95\columnwidth, height=2.15in]{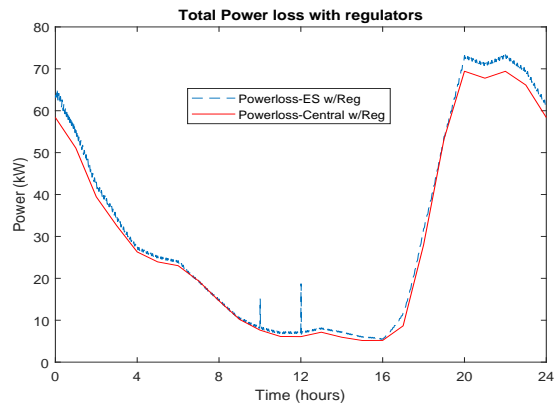}
\vspace{-0.1cm}
	\caption{Power loss comparison between centralized optimization method(red) and proposed method(blue) when regulators are activated}
\vspace{-0.1cm}
	\label{fig: PowerlossWithReg}
	\vspace{-1.20cm}
\end{figure}

It is observed that the proposed method can track optimal power loss when the regulator taps change (see Fig. \ref{fig: PowerlossWithReg}). Specifically, the proposed approach is able to track the centralized solution for optimal power loss even when legacy voltage control devices (here voltage regulators) are allowed to operate based on their local control settings.


\vspace{-0.3cm}

\section{Conclusion}
This paper presents a localized model-free Volt-VAR optimization (VVO) approach for network power loss minimization. The proposed controller is a combination of the extremum seeking algorithm to achieve the network-level objective without communicating with other decision-making agents and an adaptive droop controller to achieve a stable response under fast varying phenomena. The paper also theoretically evaluates the conditions for which the obtained localized results are close to centralized optimal power flow (OPF) solutions. For this, we investigate the convexity conditions for the problem objective (power loss minimization) with respect to the decision variables (reactive power dispatch). 
It is shown that the proposed method is capable of reducing the power losses and maintaining the bus voltage within limits in unbalanced power distribution systems. Further, the obtained power loss using the proposed controller is close to the results obtained using centralized OPF solutions. The controller is also capable of managing control instability and high variance PV generation.
	\bibliographystyle{ieeetr}
\vspace{-0.5cm}
\bibliography{references}

\balance

\end{document}